\documentclass[aps,prl,twocolumn,superscriptaddress,showpacs,floatfix,amsmath,amssymb]{revtex4-1}
\usepackage{amssymb}
\usepackage{bm}
\usepackage{url}
\usepackage{graphicx}
\usepackage{color}

\begin{document}

\title{Buoyancy-driven flow through a bed of solid particles produces a new
  form of Rayleigh-Taylor turbulence}

\author{G. Sardina}
\affiliation{Linn\'e Flow Centre and SeRC (Swedish e-Science Research Centre), KTH Mechanics, S-100 44 Stockholm, Sweden}
\affiliation{Division of Fluid Dynamics, Department of Mechanics and Maritime Sciences, Chalmers University of Technology,
Gothenburg, Sweden}
\author{L. Brandt}
\affiliation{Linn\'e Flow Centre and SeRC (Swedish e-Science Research Centre), KTH Mechanics, S-100 44 Stockholm, Sweden}
\author{G. Boffetta}
\affiliation{Dipartimento di Fisica and INFN, Universit\`a di Torino, 
via P. Giuria 1, 10125 Torino, Italy}
\author{A. Mazzino}
\affiliation{Department of Civil, Chemical, and Environmental Engineering, University of Genova and INFN, via Montallegro 1, Genova 16145, Italy}

\begin{abstract}
Rayleigh--Taylor fluid turbulence through a bed of rigid, finite-size, spheres
is investigated by means of high-resolution Direct Numerical Simulations (DNS),
fully coupling the fluid and the solid phase via a state-of-the art  Immersed
Boundary Method (IBM). The porous character of the medium reveals a totally
different physics for the mixing process when compared to the well-known
phenomenology of classical RT mixing. For sufficiently small porosity, the
growth-rate of the mixing layer is linear in time (instead of quadratical) and
the velocity fluctuations tend to saturate to a constant value (instead of
linearly growing).  We propose an effective continuum model to fully explain
these results where porosity originated by the finite-size spheres is
parameterized by a friction coefficient.
\end{abstract}


\maketitle 

\textit{Introduction.} --
Rayleigh-Taylor (RT) turbulence is strongly
influenced by physical  phenomena such as rotation
\cite{carnevale2002rotational,baldwin2015inhibition,boffetta2016rotating}
surface tension \cite{chertkov2005effects},
viscosity variations and/or viscoelastic effects \cite{boffetta2010polymer,
zhou2017rayleigh,zhou2017rayleigh2,boffetta2017incompressible}.
Little is know about RT turbulence, and buoyancy-driven turbulence in
general, in porous media. This is in spite of the importance of the problem
in a variety of  environmental applications. Among the many possible examples,
we mention
here the geological storage of $CO_2$ in saline aquifers \cite{MICHAEL2010659}
in order to mitigate the effects of emissions on climate changes, and all processes
involving the injection of a hot fluid into a cooler, fluid-saturated, subsurface rock including the so-called thermal Enhanced Oil Recovery 
\cite{Obembe2016}.\\
All these applications
have renewed the interest to understand RT-induced mixing  in porous media
\cite{HN2014,slim_2014,NAKANISHI2016224,kalisch2016rayleigh,Binda17}.
Accurate prediction of the performance of these processes requires a model 
describing  the fully-coupled dynamics  of the rock--fluid system. 
Our aim here is to propose a simple model of porous buoyancy-driven  turbulent
flow whose universal properties are extracted via a phenomenological theory.
The simple model we study here shares with all real complex systems two key
properties:
i) the fluid motion is triggered by the RT instability; ii) the 
fluid motion evolves in a porous medium. We
will show that the fluid-structure interaction problem radically changes the
classical RT scenario giving rise to new physics which can be captured
by simple theoretical arguments.
%
\begin{figure}[h]
\includegraphics[width=0.3\columnwidth]{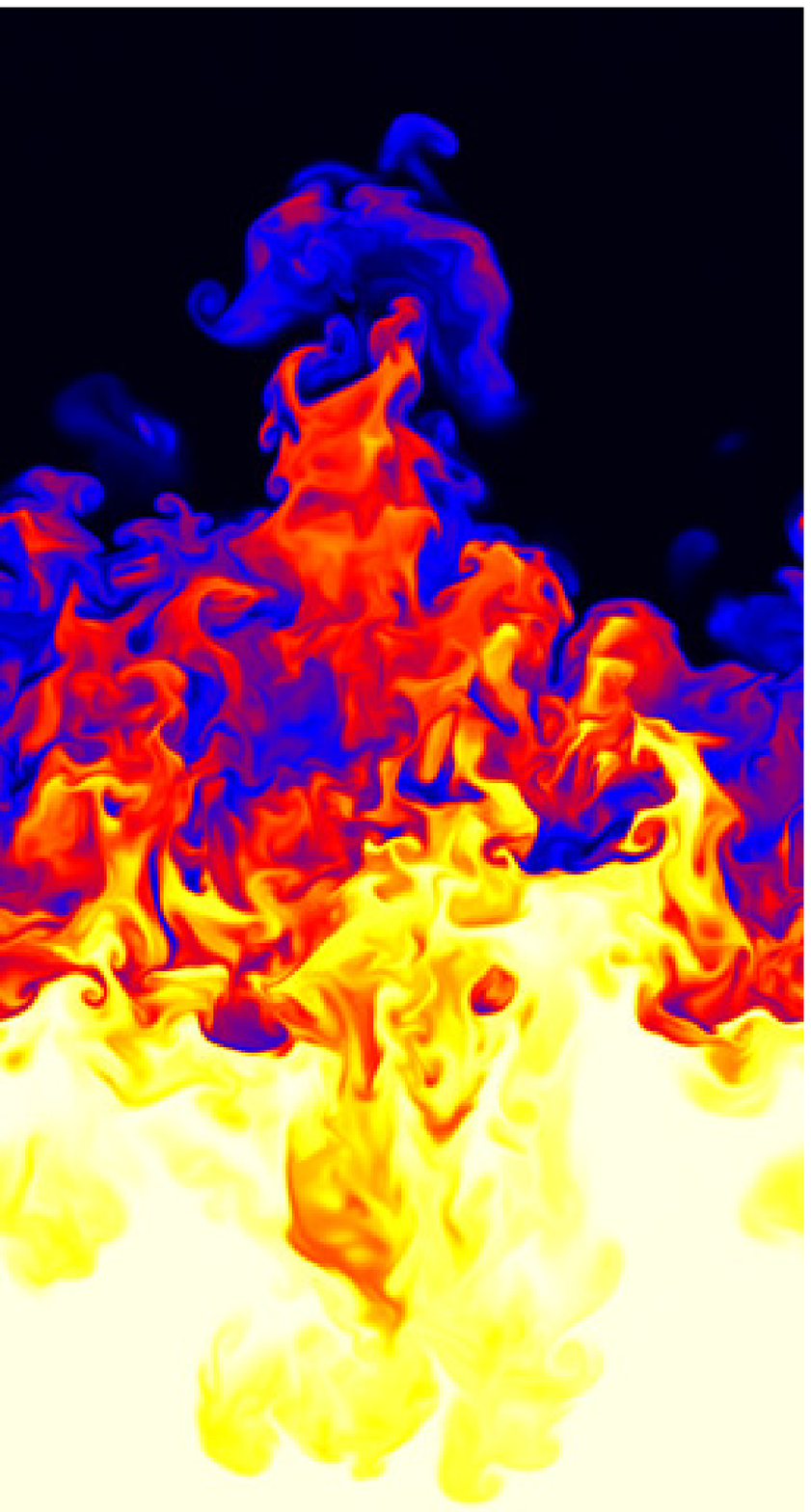}
\includegraphics[width=0.3\columnwidth]{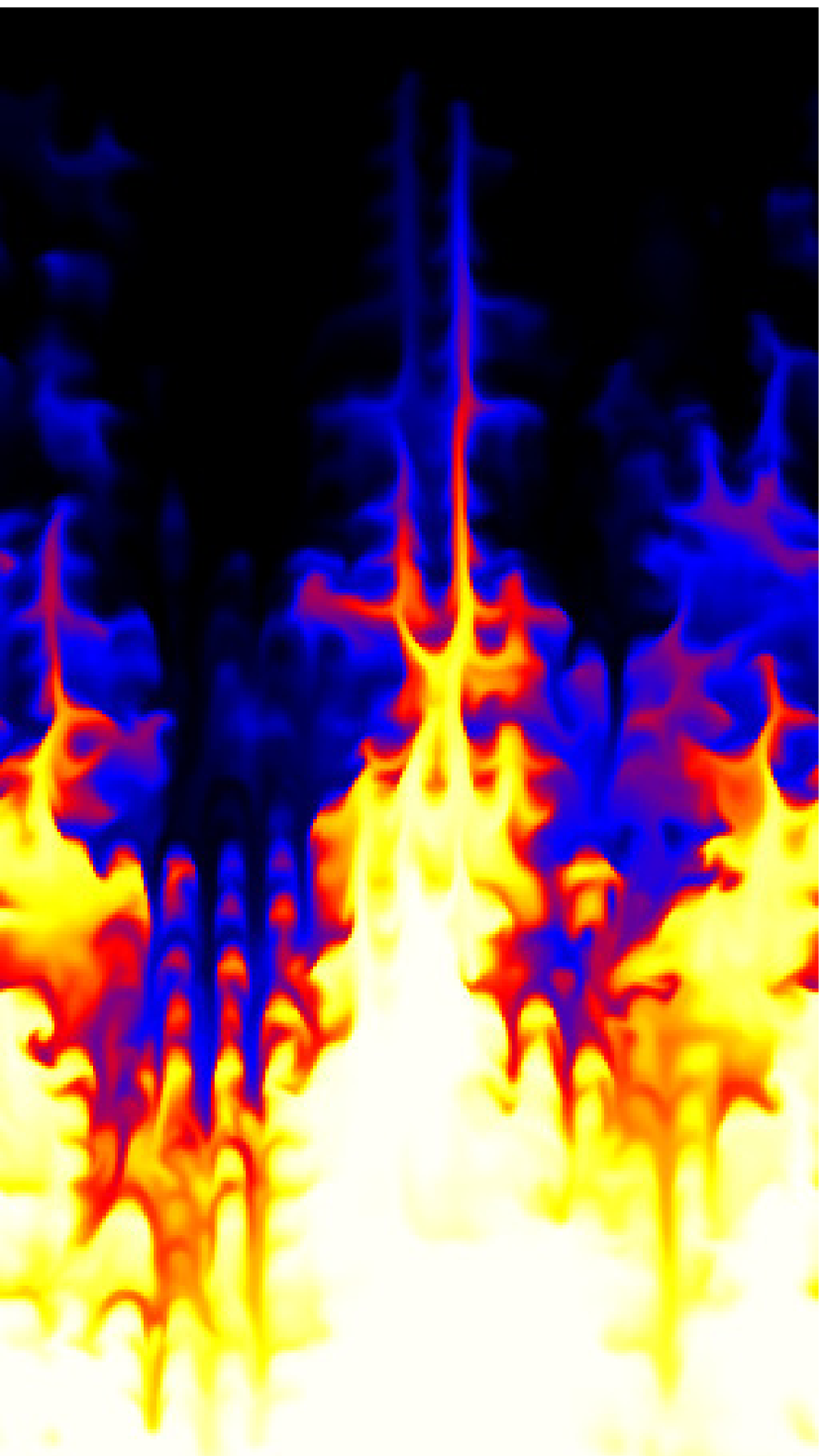}
\includegraphics[width=0.3\columnwidth]{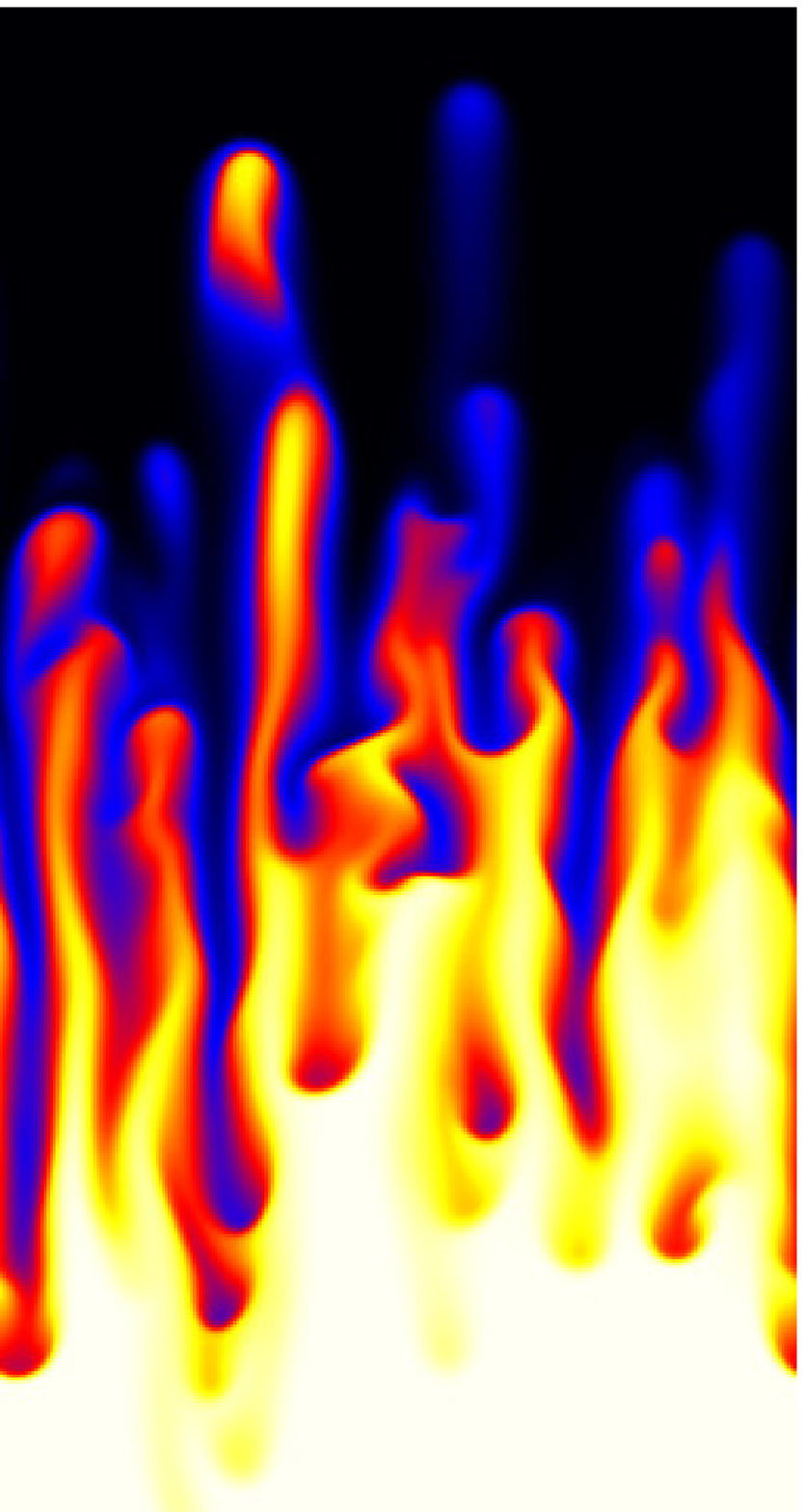}
\caption{Vertical sections of the temperature field for 
Rayleigh-Taylor turbulence. 
Left: standard RT turbulence in homogeneous fluid with 
porosity coefficient $\phi=1$.
Center: fully resolved simulation of porous RT turbulence with $\phi=0.6$.
Right: effective homogeneous model
with friction coefficient $\alpha \tau=3$.}
\label{fig1}
\end{figure}

Here, we address the problem of RT turbulence in porous media
by extensive numerical simulations of a fully resolved two-phase
flow, representing a disordered distribution of solid particles 
(spheres) in the computational domain for different values of
porosity. A state-of-the-art immersed-boundary method is employed 
to simulate the presence of the particles \cite{wpcode,Lashgari2014,Picano2015}.
We find that the growth of the mixing layer
is strongly affected by the presence of particles and,
for sufficiently large concentrations, the mixing layer grows linearly
in time.
Velocity fluctuations are reduced and saturate to a
constant value in the limit of large concentrations, with increasing
anisotropy. The presence of particles also suppresses
the turbulent heat transfer. 
The resulting phenomenology is in sharp contrast
with the well-known quadratic growth rate of the mixing layer, 
accompanied by the linear growth in time of velocity fluctuations
occurring in classical RT turbulence \cite{boffetta2017incompressible}.
Moreover, we compare the results of the fully resolved model
with an effective continuous model in which the porosity is parameterized
by a friction coefficient, a model for which simple theoretical predictions
are possible, and find a good agreement with the results of the fully 
resolved model.

\textit{Model for porous RT turbulence.} --
We consider the Boussinesq model for the buoyancy-driven incompressible flow with velocity 
${\bf u_f}({\bf x},t)$ and temperature $T({\bf x},t)$
in the presence of gravity ${\bf g}=(0,0,-g)$
\begin{equation}
{\partial {\bf u_f} \over \partial t}+ {\bf u_f} \cdot {\bf \nabla} {\bf u_f} 
= - {\bf \nabla} p + \nu \nabla^2 {\bf u_f} - \beta {\bf g} T +\bf f
\label{eq:bous_u}
\end{equation}
where $\nu$ is the kinematic viscosity of the fluid, $p$ the pressure, 
$\beta$ the thermal expansion
coefficient and ${\bf f}({\bf x},t)$ is the immersed-boundary forcing 
that accounts for the presence of the particles. 
The temperature equation is solved in all the computational domain
for both fluid and solid phases
\begin{equation}
{\partial T \over \partial t}+ {\bf u_{cp}} \cdot {\bf \nabla} T =
 \nabla \cdot ( \kappa_{cp} \nabla T)
\label{eq:bous_t}
\end{equation}
where $\bf u_{cp}$ and $\kappa_{cp}$ are the velocity and thermal
diffusivity of the combined phase. These last
two quantities can be expressed in a Volume of Fluids (VoF) formulation 
\cite{hirt81}, based on the combined single phase 
values and on the local volume fraction. 
Since in the case considered here the particles do not move,
i.e. the forcing term ${\bf f}$ does not depend on time,
${\bf u_{cp}}=(1-\xi) {\bf u_{f}}$, where $\xi({\bf x})$ is a phase 
indicator field equals to $0$ in the fluid phase and to $1$ in the 
solid phase. 
Similarly the combined thermal diffusivity is written as 
$\kappa_{cp}=(1-\xi) \kappa_{f}+\xi \kappa_s$ 
where $\kappa_{f}$ and $\kappa_s$ are the thermal diffusivity of the fluid
and solid phases respectively \cite{Mehdi2018Heat}.
The computational domain contains 
a random distribution of $N$ solid spherical particles (obstacles)
of macroscopic radius $r_p$. Particles are fixed in space and the
no-slip and no-penetration boundary conditions on their surface are 
imposed indirectly via the forcing term  ${\bf f}({\bf x})$ 
in (\ref{eq:bous_u}).
Further details on the numerical method  can be found in 
the Supplemental Material \cite{Supplementary}
and in \cite{Mehdi2018Heat,Mehdi2018_2,romaetal,kemfro}.

The velocity and temperature fields in (\ref{eq:bous_u}-\ref{eq:bous_t})
are defined in a domain of volume $V=L_x \times L_y \times L_z$,
with periodic conditions on the domain boundaries. 
The porosity of the domain, the ratio of the void volume over 
the total volume, is 
$\phi=1-N V_p/V=1-\langle \xi({\bf x}) \rangle$ where 
$V_p=(4/3) \pi r_p^3$ is the volume of a single particle
and $\langle \cdot \rangle$ represent the volume average.

We perform  direct numerical simulations of 
Eqs.~\ (\ref{eq:bous_u}-\ref{eq:bous_t})
at different values of porosity. 
For simplicity and numerical convenience, the simulations described 
in this Letter assume 
$\kappa_f=\kappa_s=\kappa_{cp}=\nu$.
The domain size has horizontal dimensions $L_x=L_y=32 r_p$ 
and vertical height $L_z=128 r_p$. A resolution of 16 points per particle 
diameter is used, giving a total of $N_x=N_y=256$ and $N_z=1024$ grid points
on a regular grid.
Numerical results are averaged over $4$ independent runs starting with
different initial perturbations and are presented as dimensionless quantities 
using $L_z$ and $\tau=(L_z/Ag)^{1/2}$ as space and time
units respectively.

The initial condition for Rayleigh-Taylor instability and
turbulence is a layer of cooler (heavier) fluid over a warmer (lighter) 
layer 
at rest, i.e.\ $T({\bf x},0)=-(\theta_0/2) sgn(z)$ 
($T=0$ is the reference temperature) and
${\bf u_f}({\bf x},0)=0$ where $\theta_0$ is the initial temperature
jump which defines the Atwood number  $A=(1/2) \beta \theta_0$.
This initial condition is unstable and after the
linear instability phase, the system develops a turbulent
mixing zone that grows in time starting from the plane $z=0$ 
\cite{boffetta2017incompressible}.

The phenomenology of the pure fluid case ($\phi=1$) is well known
\cite{boffetta2017incompressible,zhou2017rayleigh}. 
After the initial linear instability, 
the flow enters into a nonlinear phase where a turbulent mixing layer
is produced and evolves in the vertical direction. 
The mixing layer amplitude can be defined in terms of the mean 
vertical temperature profile 
$\overline{T}(z,t) \equiv {1 \over L_x L_y} \int T({\bf x},t) dx dy$ as the
region of width $h$ in which 
$|\bar{T}(z)| \le {\theta_0 \over 2} r$ where 
$r<1$ is a threshold (typically $r=0.9$).
In the turbulent phase the width of
the mixing layer grows asymptotically as $h(t)=c A g t^2$, 
while vertical and horizontal velocity fluctuations grow linearly
in time, with vertical fluctuations about two times larger than
horizontal fluctuations and isotropic velocity gradients
\cite{boffetta2010statistics}.
The determination of the dimensionless coefficient $c$ has been the
object of many numerical and experimental studies both in $3D$,
where it is in the range $0.02-0.04$ 
\cite{cabot2006reynolds,vladimirova2009self,boffetta2010nonlinear,zhou2017rayleigh}, 
and in $2D$ \cite{young2001miscible,clark2003numerical,celani2006rayleigh}.

\begin{figure}[t]
\includegraphics[width=.9\columnwidth]{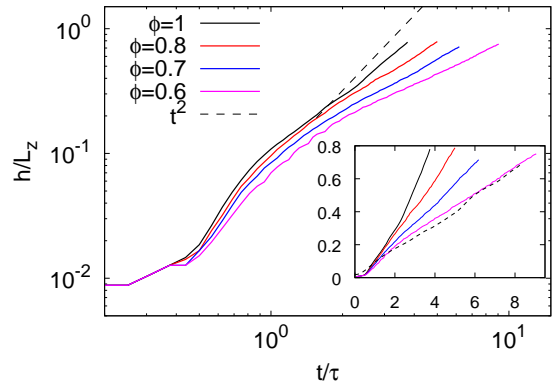}
\caption{Temporal evolution of the mixing layer $h$
in four simulations of porous RT turbulence with different values 
of the porosity $\phi$.
Dashed line represents the $t^2$ behavior. 
Inset: the same quantities in lin-lin plot to emphasize the 
linear growth at later times for the two cases at smallest porosities. 
The dashed line represents the result from the continuous model 
with $\alpha \tau=3$.}
\label{fig2}
\end{figure}

Figure~\ref{fig1} shows a section of the temperature field for 
classic RT turbulence and a case with porosity $\phi=0.6$. 
Qualitative differences between the two cases are 
evident; in particular, the presence of strongly anisotropic, vertically
elongated, plumes in the porous case. These differences are quantified in 
Fig.~\ref{fig2}, where we plot the time evolution of the mixing layer $h(t)$
for different values of the porosity, starting from the standard case $\phi=1$.
We observe that the presence of solid particles strongly reduces 
the growth of the mixing layer. 
While in the pure fluid case the mixing layer at late times
follows the classical $t^2$ 
law \cite{boffetta2017incompressible}, 
already for $\phi=0.8$ it shows a different scaling law and,
for the smallest values of porosity $\phi=0.7$ and $\phi=0.6$,
the growth becomes linear (see inset of Fig.~\ref{fig2}). 
Moreover also the coefficient of the linear growth depends on the porosity. 
We notice that at short times, $t/\tau<0.5$, the presence of the
particles has no effects on the evolution of $h(t)$ since the width of the 
mixing layer is here comparable with interparticle scale.

\begin{figure}[t]
\includegraphics[width=.9\columnwidth]{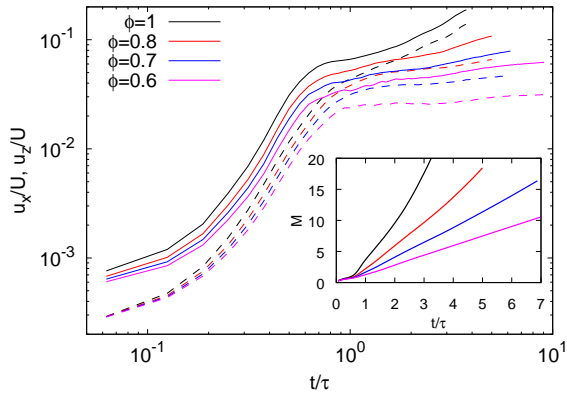}
\caption{Temporal evolution of the vertical (solid lines) and 
horizontal (dashed lines) rms velocities for the different values
of the porosity under investigation. Inset: Temporal evolution of the mixed mass ${\cal M}$ for
different values of porosity.}
\label{fig3}
\end{figure}

The reduced growth of the mixing layer is associated to the suppression
of the turbulent velocity fluctuations. In Fig.~\ref{fig3}, we show 
the horizontal $u_x(t)$ and vertical $u_z(t)$ rms velocities
in the mixing layer.
These are computed in a phase averaged sense as ${u_{x}}(t) = \langle ({\bf u_{f}}\cdot {\hat{\bf x}})^2 \rangle^{1/2} $, $\hat{\bf x}$ being the unit vector along the $x$-axis,
and similarly for $u_z$, where brackets indicate average over the 
mixing layer.
Figure~\ref{fig3} shows that both components are reduced 
in the presence of particles.
For the smallest value of porosity, the velocity fluctuations become almost 
constant at large times. This is in agreement
with the linear growth of the mixing layer observed in Fig.~\ref{fig2}.
We observe also a small increment of the anisotropy of the velocity 
components $u_z/u_x$ with respect to the case of pure fluid $\phi=1$,
which is not surprising given the elongated structures observed in 
Fig.~\ref{fig1}.

The growth of the mixing layer is a basic measurement of the amount
of mass mixed by the turbulent flow. Recently, a more direct 
indicator of the mixed mass, ${\cal M}$, has been introduced
which has the advantage of being a conserved inviscid quantity
\cite{zhou2016asymptotic}.
It is defined by the integral 
\begin{equation}
{\cal M}=\int 4 \rho Y_1 Y_2 d^3 x
\label{eq:mixed}
\end{equation}
where $\rho$ is the mixture density and the mass fractions, in the present case of a symmetric
temperature jump, are $Y_1({\bf x})=(\theta_0/2-T)/\theta_0$
and $Y_2({\bf x})=(\theta_0/2+T)/\theta_0$.

While for the higher values of porosity ${\cal M}$ follows
the $t^2$ behavior observed in the standard RT turbulence 
\cite{zhou2016asymptotic}, for the lower values $\phi=0.7$ and 
$\phi=0.6$ it displays a clear linear behavior (see inset of Fig.\ \ref{fig3}). 

\begin{figure}[t]
\includegraphics[width=.9\columnwidth]{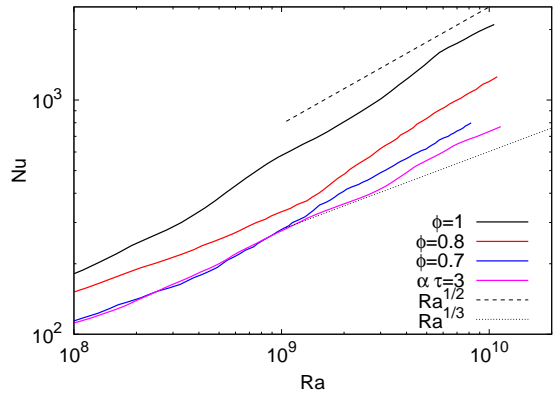}
\caption{Nusselt number $Nu$ as a function of the Rayleigh number $Ra$
from three simulations of porous RT turbulence and from one 
simulation of the continuous model. From top to bottom: $\phi=1$ (black line),
$\phi=0.8$ (red line), $\phi=0.7$ (blue line) $\alpha \tau=3$ (pink line).
The upper dashed line represents the ultimate state scaling 
$Ra^{1/2}$, the lower dotted line is the $Ra^{1/3}$ scaling.}
\label{fig4}
\end{figure}

Figure~\ref{fig4} shows the dimensionless turbulent heat transfer 
$ Nu=1+\langle {\bf u_{f}}\cdot {\hat{\bf z}}\,  T \rangle h/(\kappa_{f} \theta_0)$
as  a function of the Rayleigh number, defined for RT turbulence 
as  $Ra=A g h^3/( \nu \kappa_{f} )$. 
We observe large fluctuations for all the values of porosity,
even after averaging over realizations. Nonetheless it is possible 
to observe a reduction of $Nu$, for given $Ra$, by decreasing the
value of porosity. 
A similar behavior has been observed
in the case of rotating Rayleigh-Taylor turbulence where the reduction
of $Nu$ is produced to the decoupling of velocity and temperature fluctuations
due to the bi-dimensionalization of the flow \cite{boffetta2016rotating}.
For large porosity, $\phi=1$ and $\phi=0.8$, the scaling is in 
agreement with the so-called {\it ultimate state} regime $Nu \simeq Ra^{1/2}$
already observed in the pure fluid case
\cite{boffetta2017incompressible}.
For smaller values of $\phi$ there is a clear indication of a transition 
to a different regime compatible with a
$Ra^{1/3}$ scaling. 
Indeed, assuming that $h \sim t$ and $u_{rms} \sim t^0$ we obtain $Nu \sim t$ and $Ra \sim t^3$ which imply $Nu \sim Ra^{1/3}$.


\textit{Interpretation in terms of an effective model} --
Let us now show that the features of the porous RT turbulence 
can be obtained by an effective continuous model (without particles), 
in which the porous medium is parameterized by a friction coefficient.
The model is obtained by averaging the microscopic equation over a 
volume which includes many particles and therefore filters 
the discrete nature of the porous medium. 
In the limit of small particles, the porous medium is considered 
as an homogeneous fluid with an additional
effective friction term $-\alpha {\bf u}$ added to the momentum equation
(\ref{eq:bous_u}) \cite{breugem2006influence}.
The friction coefficient $\alpha$ is:
\begin{equation}
\alpha = \nu {45 (1-\phi)^2 \over r_p^2 \phi^2} \, .
\label{eq:alpha}
\end{equation}
We remark that the use of a continuous model for the problem discussed 
in this Letter is not justified a priori, since there is no large-scale separation between particles size and box size. Moreover,
particles are not very small to guarantee the presence of a Stokes
flow in the pores.  This is why the fluid inertia contribution is retained.
The continuous model can be corrected taking into account 
finite particle Reynolds number $Re_p=r_p u /\nu$ (where $u$ represents
the magnitude of the flow velocity around the particle) by the factor
$\left(1+{\phi \over 50 (1-\phi)} Re_p \right)$ \cite{breugem2006influence}.
For simplicity, in the following we consider the extension of the continuous
model
to the Boussinesq equations in the limit of small particles with 
linear friction (\ref{eq:alpha}) only and we find that it
is able to reproduces many of the results of the full microscopic
model and sheds light upon the mechanism at the basis of the results
discussed in the previous Section.

In the limit of large porosity, $\phi \simeq 1$,
the friction coefficient (\ref{eq:alpha}) vanishes and therefore
we expect that the standard RT turbulence phenomenology holds.
Therefore in this limit we can assume that 
$h \simeq \beta g \theta_0 t^2$ 
and $U \simeq \beta g \theta_0 t$.
On dimensional grounds, by using these scaling laws, 
one sees that $\alpha {\bf u}$
becomes dominant over ${\bf u} \cdot {\bf \nabla u}$ 
in (\ref{eq:bous_u}) after a time $t_{\alpha} \simeq 1/\alpha$. 
Therefore, for $t > t_{\alpha}$ we expect a different phenomenology 
given by the balance of the buoyancy term which injects energy and the 
friction term which removes the energy in the system. This 
balance gives the new scaling laws
\begin{eqnarray}
h &\simeq& {\beta g \theta_0 \over \alpha} t  \, ,
\label{eq:h_friction} \\
U &\simeq& {\beta g \theta_0 \over \alpha}  \, .
\label{eq:u_friction}
\end{eqnarray}
Therefore, already at the level of dimensional analysis, the effective model
is able to reproduce the behavior observed in the fully microscopic model,
i.e. the saturation of velocity fluctuations and the linear growth of 
the mixing layer.

\begin{figure}[t]
\includegraphics[width=.9\columnwidth]{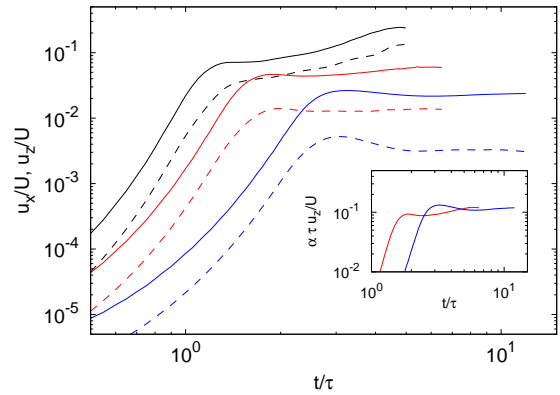}
\caption{Temporal evolution of the vertical (solid lines) and horizontal
(dashed lines) rms velocities from three simulations of effective RT 
turbulence with friction coefficient $\alpha \tau=0$ (black lines), 
$\alpha \tau=3$ (red lines) and $\alpha \tau=5$ (blue lines).
Inset: vertical rms velocities multiplied by the friction coefficient
for the cases $\alpha \tau=3$ and $\alpha \tau=5$.}
\label{fig5}
\end{figure}

In Fig.~\ref{fig5} we plot the time evolution of the rms of the 
horizontal and vertical velocities for three different simulations
of the effective model: one for the standard RT with $\alpha=0$ and 
two with larger  values of the friction coefficient. 
The case $\alpha \tau=3$ corresponds to the case
$\phi=0.6$ according to (\ref{eq:alpha}) and will be used
to make a quantitative comparison of the homogeneous model with the
full microscopic model.
As in the microscopic model, we observe that
while for $\alpha=0$ the large-scale velocity
grows linearly in time (after an initial transient), 
in the 
simulations with friction the velocity saturates
to a constant value. Moreover, anisotropy increases with $\alpha$,
as the horizontal velocity is suppressed more than the vertical one,
a feature also observed in the microscopic model (see Fig.~\ref{fig3}).
In the inset, we report the vertical rms velocity multiplied by 
the friction coefficient $\alpha$, which, according to (\ref{eq:u_friction}),
gives a constant value independent on $\alpha$.

\begin{figure}[h]
\includegraphics[width=.9\columnwidth]{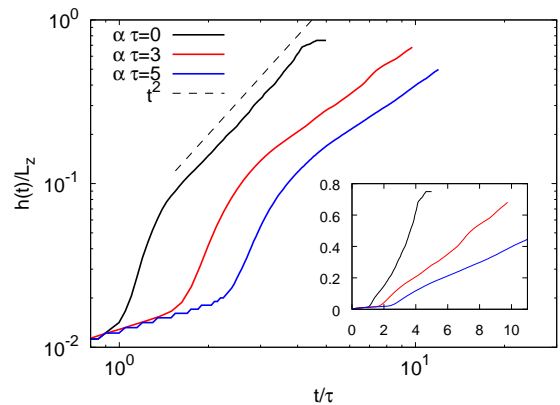}
\caption{Temporal evolution of the mixing layer width from the continuous effective model
for the three cases with $\alpha \tau=0$ (black), 
$\alpha \tau=3$ (red) and $\alpha \tau=5$ (blue).
Dashed line represents the $t^2$ law. Inset: the same quantities
in lin-lin plot.}
\label{fig6}
\end{figure}

Figure~\ref{fig6} shows the evolution of the mixing layer width,
$h(t)$, from the three simulations of the effective model at increasing
values of the friction coefficient.
The presence of friction slows down the growth of the mixing 
layer but also changes its slope. For the largest values of 
$\alpha$, the growth becomes linear (see inset of Fig.~\ref{fig6})
in agreement with the prediction (\ref{eq:h_friction}).

The case $\alpha \tau=3$ is also plotted in the inset of 
Fig.~\ref{fig2} for a direct comparison with the full model. 
It is evident that the simple homogeneous model is able to reproduce
quantitatively the behavior and the transition observed in the 
microscopic model.

\textit{Conclusions.} --
We have numerically studied Rayleigh-Taylor turbulence in the presence of fixed macroscopic solid particles, 
for different values of porosity coefficient. We have shown that
the presence of particles reduces the growth of the mixing layer,
which for small porosity follows asymptotically a linear 
behavior. In this regime turbulent velocity fluctuations saturate
to a constant value. We have interpreted these results in terms
of a continuous homogeneous model with an additional linear 
friction term representing the effective porosity of the medium.
Dimensional analysis predicts that the friction term at late times
modifies the asymptotic growth of the mixing layer. This 
is confirmed by extensive simulations of the effective model which
is shown to reproduce the main features observed in the full model.

\noindent {\sc Acknowledgments.} 
G.B. acknowledges financial support by the project CSTO162330 
{\it Extreme Events in Turbulent Convection} and from
the {\it Departments of Excellence} grant (MIUR).
HPC center CINECA is gratefully acknowledged for computing resources.



\bibliography{biblio}

\end{document}